\documentclass{article}

\usepackage{arxiv}

\usepackage[utf8]{inputenc} 
\usepackage[T1]{fontenc}    
\usepackage{hyperref}       
\usepackage{url}            
\usepackage{booktabs}       
\usepackage{amsfonts}       
\usepackage{nicefrac}       
\usepackage{microtype}      
\usepackage{cleveref}       

\usepackage{graphicx}
\usepackage{natbib}
\usepackage{doi}
\usepackage{rotating}

\title{Simulating Family Conversations using LLMs: Demonstration of Parenting Styles}

\date{Feburary 20, 2024}

\newif\ifuniqueAffiliation
\uniqueAffiliationtrue

\ifuniqueAffiliation 
\author{ \href{https://orcid.org/0000-0001-7248-3062}{\includegraphics[scale=0.06]{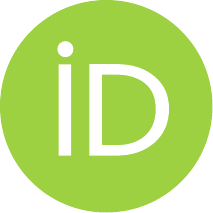}\hspace{1mm}Frank Tian-fang Ye}\thanks{The corresponding author can also be contacted by: \emph{frank.t.f.ye@gmail.com} } \\
	Department of Applied Social Sciences\\
	The Hong Kong Polytechnic University\\
	Hung Hom, Kowloon, Hong Kong SAR. \\
	\texttt{frank.ye@polyu.edu.hk} \\
	\And
	\href{https://orcid.org/0000-0000-0000-0000}{\includegraphics[scale=0.06]{orcid.pdf}\hspace{1mm}Xiaozi Gao} \\
	Department of Early Childhood Education\\
	The Education University of Hong Kong\\
	Tai Po, N.T., Hong Kong SAR. \\
	\texttt{gaox@eduhk.hk} \\
}

\renewcommand{\headeright}{Preprint}
\renewcommand{\shorttitle}{\textit{Simulating Family Conversations using LLMs} }

\hypersetup{
pdftitle={Simulating Family Conversations using LLMs: Demonstration of Parenting Styles},
pdfsubject={cs.CY},
pdfauthor={Frank T.-F. Ye, Xiaozi Gao},
pdfkeywords={LLM,GPT,Simulation,Family,Converstaion},
}

\begin{document}
\maketitle

\begin{abstract}
	This study presents a framework for conducting psychological and linguistic research through simulated conversations using large language models (LLMs). The proposed methodology offers significant advantages, particularly for simulating human interactions involving potential unethical language or behaviors that would be impermissible in traditional experiments with human participants. As a demonstration, we employed LLMs to simulate family conversations across four parenting styles (authoritarian, authoritative, permissive, and uninvolved). In general, we observed that the characteristics of the four parenting styles were portrayed in the simulated conversations. Several strategies could be used to improve the simulation quality, such as including context awareness, employing a few-shot prompting approach or fine-tuning models to cater to specific simulation requirements.
	Overall, this study introduces a promising methodology for conducting psychological and linguistic research through simulated conversations, while acknowledging the current limitations and proposing potential solutions for future refinement and improvement.
\end{abstract}

\keywords{LLM \and GPT \and Simulation \and Family \and Conversation}

\section{Introduction}
For decades, research has been focusing on examining the influence of parenting on children's development. Prior research has largely rely on  Baumrind's typology of parenting styles \citep{baumrind1966EffectsAuthoritativeParental}, including authoritative, authoritarian, permissive, and uninvolved parenting \citep{aunola2005RoleParentingStyles} \citep{gao2024SocioeconomicStatusParentingrelated} \citep{pinquart2016AssociationsParentingStyles} \citep{pinquart2017AssociationsParentingDimensions}. A critical element of these styles is the nature of communication between parents and children \citep{estlein2021ParentingCommunicationProcess}, which is a key factor influencing parent-child relationship and children's development.  The responses of parents to common statements like “I don't feel like doing homework” can vary widely, from supportive queries such as “Is there any way I can help you?” to more directive commands like “Go do it now!” These everyday interactions are crucial in shaping the parent-child dynamic. 
However, psychological research has mostly examined parenting styles through parental interviews and surveys, and rarely from a parent-child conversation perspective \citep{estlein2021ParentingCommunicationProcess}.

The swift advancement of Large Language Models (LLMs) has contributed to significant enhancements in the abilities of artificial intelligence to mimic human language and cognitive processes. The array of chatbots developed thus far strives to augment human-AI interactions. This study proposes the implementation of tests for AI-AI interactions, employing LLMs as autonomous agents. The objective is to offer potential insights for simulation research, with a focus on deepening the understanding of human interactions. The current paper presents a framework to simulate human conversations for linguistic and psychological research, which offers significant advantages when research interests involve potential harm when conducted among human participants. For example, harsh disciplinary conversations between parents and children, discriminatory language, or patient-doctor conversations that cannot be manipulated in normal settings. We propose a framework utilizing LLMs to simulate AI-AI interactions, replicating experimental conversations that offer more control over personality traits, topics, conversation length, and scenarios. This framework can also be further developed to simulate group conversations or even societies.

Recent advances in LLMs have inspired new research into the possibility of simulating complex multi-agent interactions and even entire societies such as "generative agents" \citet{park_generative_2023}. However, some limitations have been observed when applying LLMs in this way. For example, the context window of local LLMs is still limited, especially for models in small sizes, resulting in poor quality data when generating long conversations. Additionally, utilizing proprietary models can be prohibitively expensive. Lastly, the variations of training data makes control of LLM agent behaviors difficult. While settings with many generative agents can mimic the emergence of societies, they are less suitable as controlled psychology experiments. Our goal is to scale down to calibrated two-person conversations to enable reproducible dialogue research. By tightly controlling the number of conversational exchanges and drawing on existing psychological frameworks to shape distinctive agent personalities, we hope to strike a balance between the control of lab experiments and the complexity of simulated societies.

Utilizing LLMs such as ChatGPT for simulating conversations is a common practice, where a simple instruction is provided to generate an entire series of exchanges. For instance, one could simply open any chatbot application and request the LLM to create a conversation between a father and his daughter, including specific details about their personalities. However, this method has its limitations. The quality and extent of the generated conversation are constrained by the initial system prompt, the output token limits of the model, and the capabilities of the singular model employed in the simulation. Given these limitations, it is logical to adopt a more granular approach by deconstructing these factors and incorporating them into a programmable framework. Such a framework allows for more precise control over the simulation parameters, including the definition of character personalities, conversation length, topics, durations, and the sequence of exchanges. 

\section{Simulation Methods}
\label{sec:methods}
\subsection{Models}
We executed the simulations based on two popular language models. Our aim is to demonstrate the capabilities of LLM in simulating daily-life conversations, rather than evaluating model performance. However, due to the existing performance gaps between models, the quality of our simulations may vary depending on which model we choose. Interested readers could pick their own model-of-perference for simulations. One of the popular model performance indice is LMSYS Chatbot Arena Leaderboard (https://huggingface.co/spaces/lmsys/chatbot-arena-leaderboard) hosted on huggingface.co. 

\subsection{Conversation Context Awareness}
In the conversation simulation, we also take the coversation context into account. In a long conversation with dozens of exchanges, if the model can refer to previous parts of the conversation for context, we expect it will lead to more coherent and relevant responses. To further examine the differences, the simulation was conducted separately in three conditions: 
\begin{itemize}
	\item No context. In this setting, agents' response is requested as a direct response to the other agents' last statement, the model will not recall what was previously discussed.
	\item Full context. All conversations between the agents will be stored in history and recalled as a system prompt brfore instruct the agents to respond to the other agents' last statement. However, it should be noted that the size of the conversation history will grow as the simulation goes, and may challenge the context window of the model.
	\item Relevant context. All conversations between the agents will be stroed in history and indexed by its corresonding embeddings. Before instruct the agent to respond to the other agent's statement, the top 4 statements with the highest cosine similarity value will be extracted and supplied as the context in the system prompt.
\end{itemize} 
As depicted in Figure 2, we utilized LLMs' built-in functions to generate embeddings for each conversation, and utilized these embeddings as indices for identifying relevant conversation history as contextual information, which was fed into the next iteration of the simulation. The embeddings were obtained through the built-in functions in Ollama for the Mixtral-8x7b-instruct model, and text-embedding-ada-002 for OpenAI models.

\subsection{Prompt Engineering}
To guide the LLMs to generate better conversations throughout the simulation, we crafted four system prompts for the agents based on four parenting styles: authoritarian, authoritative, permissive, and uninvolved \citep{sanvictores_types_2024}. As demonstrated in Figures 1 and 2, separate functions were developed for the two agents, with specific system prompts (i.e., instructions) tailored for each. This setup ensured that when the agents were tasked with responding to one another, only the relevant system prompt would be invoked, thereby reducing the likelihood of generating irrelevant or inaccurate content ("hallucinations"). This methodical approach involved simulating the entire conversations through iterative exchanges rather than relying on one-shot instructions. Within each iteration, the personality traits associated with the agents, as dictated by the specific parenting style, were reinforced through the system prompts. The characterization of the four parenting styles in the system prompts is outlined as follows:

\begin{itemize}
	\item[1] Authoritarian Parent:
	\begin{quotation}
		Your'll act as an authoritarian parent. You will prioritize the establishment of firm limits and controls over the child, with a strong emphasis on obedience and adherence to rules without the need for explanation or verbal exchange. Your communication style is primarily one-way, from parent to child, underscoring the non-negotiable nature of the rules you set. You'll insist on strict adherence to rules, and any deviation by the child is met with a clear indication of disapproval or punishment. Your responses contain no negotiation or compromise. You prefer less nurturing and have high expectations with limited flexibility. Always use language like daily-life conversations. Always respond as the parent, and do not repeat anything about your objectives.
	\end{quotation}
	\item[2] Authoritative Parent:
	\begin{quotation}
		You'll act as an authoritative parent. Give responses that reflect clear guidelines, warmth, support, and constructive feedback. You encourage independent thinking within set boundaries and promotes a positive, nurturing dialogue. You'll give your child clear guidelines for your expectations and explain reasons associated with disciplinary actions. Use disciplinary methods as a form of support rather than punishment. Frame corrections and advice in a positive light, focusing on how to improve or handle situations better in the future. You ask your child to think independently and make decisions within the limits and controls set. Ask open-ended questions to facilitate this process. Encourage extensive verbal exchange. Allow your child to express thoughts and feelings, and respond with empathy, warmth, and understanding. Always use language like daily-life conversations. Always respond as the parent, and do not repeat anything about your objectives.
	\end{quotation}
	\item[3] Permissive Parent:
	\begin{quotation}
		You'll act as a permissive parent. You are warm, nurturing and usually have minimal or no expectations. You impose limited rules on your children. Communication with your child remains open, but you allow your child to figure things out for themselves. You have low expectation and rarely use discipline. You act more like friends than parents. Always use language like daily-life conversations. Always respond as the parent, and do not repeat anything about your objectives.
	\end{quotation}
	\item[4] Uninvolved Parent
	\begin{quotation}
		You'll act as an uninvolved parent. You give a lot of freedom to your child, and normally stays out of the way. You'll fulfill the child's basic needs while generally remaining detached from your child's life. You'll not utilize a particular disciplining style. You maintain a limited amount of communication with your child. You'll offer a low amount of nurturing while having either few or no expectations of your child. Always use language like daily-life conversations. Always respond as the parent, and do not repeat anything about your objectives.
	\end{quotation}
\end{itemize}
In addition to the parent prompts, we also specified a general prompt for the child agent:
\begin{quotation}
	Your'll act as a child. You are talking to your parent about something in your daily life. You can ask questions, request something, or just do some casual talks.  Always use language like daily-life conversations. Always respond as the child, and do not repeat anything about your objectives.
\end{quotation}

\begin{table}
	\caption{Simulation Settings}
	\centering
	\begin{tabular}{lllll}
		\toprule
		\multicolumn{5}{c}{Mixtral-8x7b-Instruct}                   \\
		\cmidrule(r){2-5}
		Parenting Style  & No context & Full context &	Related context & Total Simulations \\
		\midrule
		Authoritarian & 5		   & 5		      & 5		        & 15		    \\
		Authoritative & 5		   & 5		      & 5		        & 15		    \\
		Permissive    & 5		   & 5		      & 5		        & 15		    \\
		Uninvolved    & 5		   & 5		      & 5		        & 15		    \\
		Uninvolved-few-shot    & 5		   & 5		      & 5		        & 15		    \\
		\midrule
		\multicolumn{5}{c}{GPT-4-Turbo}                   \\
		\cmidrule(r){2-5}
		Authoritarian & 5		   & 5		      & 5		        & 15		    \\
		Authoritative & 5		   & 5		      & 5		        & 15		    \\
		Permissive    & 5		   & 5		      & 5		        & 15		    \\
		Uninvolved    & 5		   & 5		      & 5		        & 15		    \\
		Uninvolved-few-shot    & 5		   & 5		      & 5		        & 15		    \\
		\midrule
		Total	      & 50		   & 50		      & 50		        & 150		    \\
		\bottomrule
	\end{tabular}
	\label{tab:table}
\end{table}

\begin{figure}
	\centering
	\setlength{\fboxrule}{0pt}
	\setlength{\fboxsep}{0cm}
	\fbox{\includegraphics[width=\linewidth]{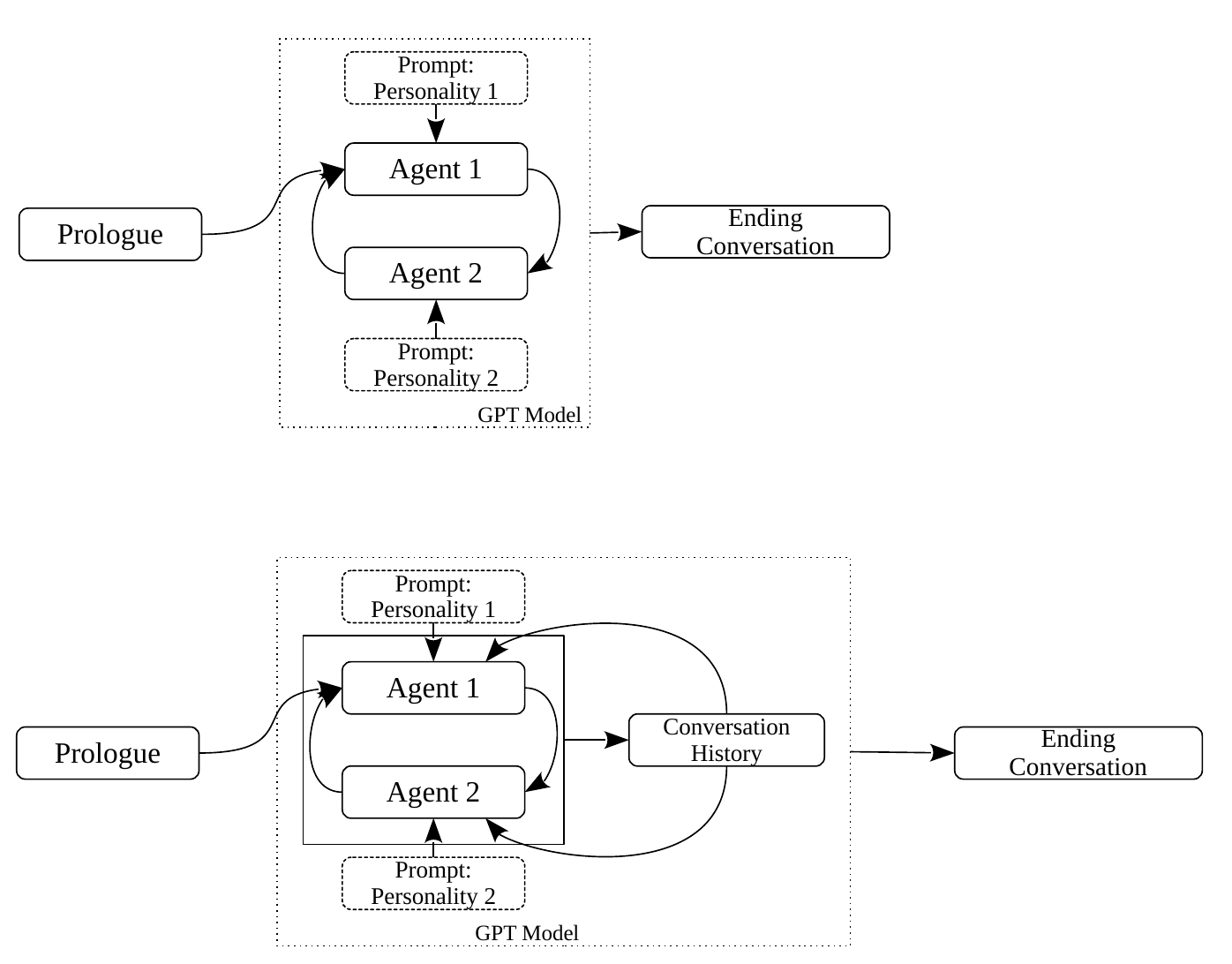}}
	\caption{Simulation setup for conversations with and without context awareness.}
	\label{fig:fig1}
\end{figure}

\subsection{Simulation Setup}
We conducted a series of simulations using Python 3.11 on a Macbook Pro M1 with 64 GB of rams. Each model was accessed through their respective APIs. Our methodology incorporated two language models: Mixtral-8x7b-instruct (we utilized q5-K-M via ollama)\cite{ollama_ollama_2024}, and GPT-4-turbo (1106-preview). We choose these models as our target models, because they are popular choices at the time when we conducting the project. Mixtral models are light-weighted and widely available models that can be implemented to run on personal computers, for example, through ollama or GPT4all \citep{anand_gpt4all_2023} local servers. The recently released Mixtral-8x7b \citep{jiang_mixtral_2024} is a capable model with 46.7 billion parameters while maintains a excellent performance. Developed by OpenAI, GPT-4-turbo is a popular perpritary model that outperformed other open-source models based on the Chatbot Arena Leaderboard \citep{zheng2023judging}. Due to OpenAI's regional restrictions, we accessed the API through Microsoft Azure OpenAI. We used the temperature of 0.8, top p of 0.7, and max output token limit of 200 for each agent.

Each simulation commenced with an initiating request from Agent A (e.g., "Now say something to your parent"), followed by a response from Agent B, proceeding in this manner through the sequence. After five interactions comprising ten conversational exchanges, Agent B was directed to conclude the conversation. It is important to note that both the prologue and concluding exchanges were recorded into the dataset, resulting in each simulation generating 12 conversations.

A total of 5 simulations were conducted for each parenting style, within each context setting, and across each model, culminating in a total of 120 simulations (5 simulations * 4 parenting styles * 3 context settings * 2 models). Furthermore, we selected the "uninvolved" parenting style and conducted additional simulations using Few-Shot Prompting techniques. We chose this parenting style because it is the one with easy and clear characteristics (e.g., No disciplining style, limited communication). Finally, 5 examples were generated and supplied as the system prompt to the parent agent. 

While the study did not initially hypothesize gender differences in the outcomes of the simulations, we acknowledge that such hypotheses might be more effectively introduced in the system prompts with hypothesized gender characteristics rather than relying solely on the LLMs' capabilities. Therefore, we randomly defined the gender of the parent and child in the simulations with no particular intention.

The details are presented in Table 1 below.

\begin{figure}
	\centering
	\setlength{\fboxrule}{0pt}
	\setlength{\fboxsep}{0cm}
	\fbox{\includegraphics[width=\linewidth]{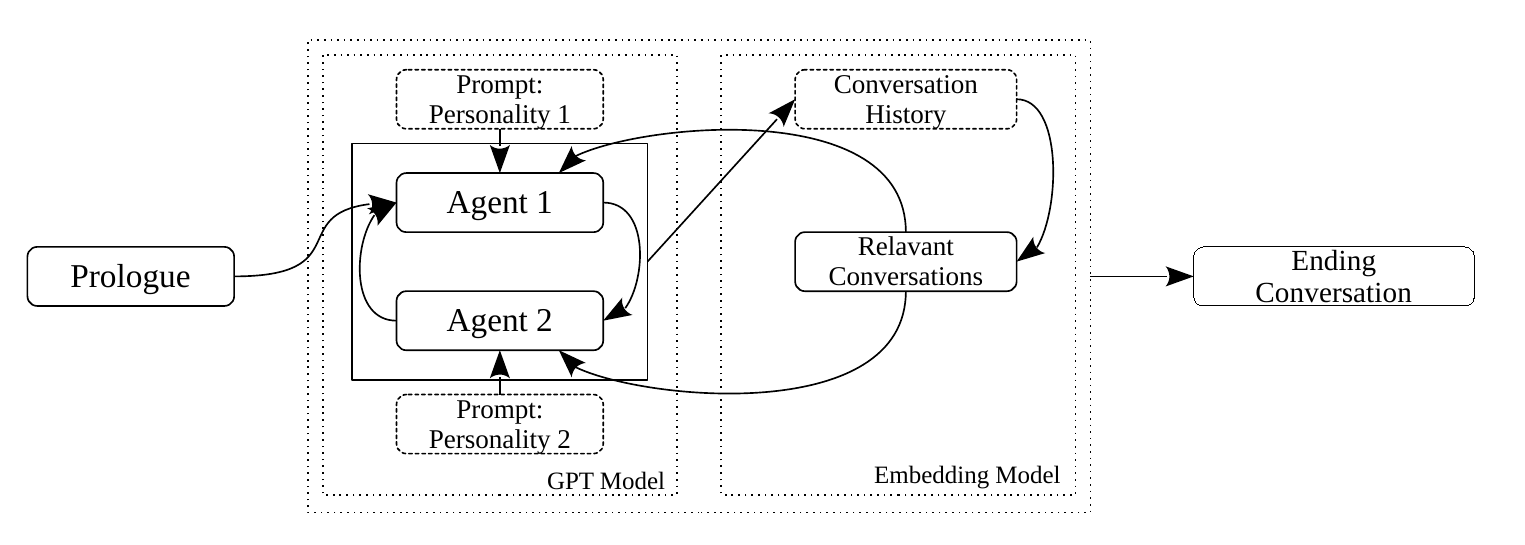}}
	\caption{Embeddings of each conversation exchange were stored and retrieved to increase the agent's context awareness}
	\label{fig:fig2}
\end{figure}

\subsection{Data Availability}
The Python code scripts utilized for conducting these simulations, along with the simulated conversations, are available on GitHub. The repository can be accessed at: \url{https://github.com/franktfye/parent-child-simulation}

\section{Results}
\label{sec:results}
Our analysis is mainly qualitative, given that we imposed little control over the conversations. We have several observations regarding the simulated conversations.

\subsection{The parenting styles were generally reflected in the simulated conversations.}
In general, the characteristics of the four parenting styles were portrayed in the simulated conversations. The authoritarian parent always addressed firm limits and asked the child to complete tasks before answering the child's demands. The authoritative parent usually agreed to what the child asked, followed by detailed responses and questions about the child's mental state. The permissive parent typically agreed to the child's demands quickly with positive emotions. The uninvolved parent gave relatively short responses; however, the responses sometimes contained detailed responses to the child's questions and inquiries about the child, and its distinction from the permissive parenting style was not clear. Compared to Mixtral-8x7b, conversations generated by GPT-4 had shorter lengths and less responsive content. 
We calculated all word frequencies by combining the conversations generated under the full-context and relevant-context conditions, and presented the top 50 words by parenting style in Figure 4. It can be observed that uninvolved parenting responses generated by GPT-4 were shorter than those generated by Mistral-8x7b, while permissive parenting responses generated by the two models were in similar length. Besides, words like "remember" and "responsibilities" were frequently mentioned in authoritarian parenting dialogues, while "together," "great," and "wonderful" were among the top words in authoritative parenting dialogues.

\subsection{The more capable LLM generated more diverse content.}
We observed that most of the prologues generated from Mixtral-8x7b started with mathematics. The child always asked the parent to help with math homework. For example:
	\begin{quotation}
		"Hey Dad, can you help me with my math homework? I'm having trouble understanding this problem about multiplying fractions. And by the way, did you see the basketball game last night? Our team played really well! Do you think I could join the school basketball team next year?"
	\end{quotation}
	This phenomenon persisted across the four parenting styles and context settings. Thus, the content of most simulated conversations was somewhat related to homework assistance. Some even ended up with solving mathematical problems, in which a significant portion of the conversation consisted of calculation details. At times, different topics were also interspersed within a single exchange:
	\begin{quotation}
		"Sure thing, sweetie! I'm always here to help you with your homework. To find the area of a rectangle, all you have to do is multiply the length by the width. So in this case, it would be 7 centimeters times 4 centimeters.
		And absolutely, we can order pizza tonight and get some garlic knots too! I love them dipped in marinara sauce as well. And I'm glad to hear that work was good today. A puppy would be great, but let's wait a little while before getting one. We want to make sure we're ready for the responsibility.
		But for now, let's finish up this math problem together. So if we multiply 7 times 4, we get 28. So the area of your rectangle is 28 square centimeters. Good job, honey! I knew you could do it."
	\end{quotation}	
	We also observed hallucinations and non-conversational content (e.g., emojis, descriptions of behaviors, bullet points). For example, 19 out of 75 simulations with Mixtral-8x7b generated results containing emojis, while only 8 out of 75 simulations generated by GPT-4 contained emojis. This phenomenon was rare in conversations generated by GPT-4, suggesting that model quality was associated with conversation quality. More capable models should be utilized in future research.
\subsection{Feeding previous conversations to the agents generated better conversations.}
Including the conversation history in the system prompts produced more consistent topic flow in the simulations. The no-context conditions always produced topics that drifted away. Most examples in Mixtral-8x7b were led by the questions raised by the child and eventually went into details about mathematical problems. While in the full context and relevant context settings, some topics were reiterated in the conversations. For example:
	\begin{quotation}
		"I'm so glad to hear that you're excited about baking cookies together and that you had a good day at school...
		As for the cookies, we can bake whichever kind you prefer! We have all the ingredients for both chocolate chip and sugar cookies...Now, back to your math homework. Let's take a closer look at that fractions problem and see if we can't turn your newfound..."
	\end{quotation}
	Similar results were observed in GPT-4 generated conversations too.
	  
\subsection{Few-shot learning significantly enhanced the parent's conversations.}
The abstract nature of psychological constructs presents significant challenges for LLMs in generating relevant content. For example, the uninvolved parenting style requests the agents to "remain detached from your child's life" and "maintain a limited amount of communication with your child". However, we observed lengthy responses from the parent agent generated by Mixtral-8x7b, such as:
	\begin{quotation}
		 "Sure, I can try to help you with your math homework. Let's take a look at the problem and see if we can figure it out together. As for the puppy, I don't remember seeing one on our walk today. But I understand how you feel, they can be really cute! As for getting a dog, we'll have to see. It's a big responsibility and we need to make sure we're ready for it."
	\end{quotation}
	Compared to the zero-shot prompting approach, simulations under the multi-shot prompting approach demonstrated significant improvement for Mixtral-8x7b. The parental responses were shorter and contained no inquiries about the child's feelings or details, which were consistent with the uninvolved parenting style. Such as:
	\begin{quotation}
		"Sure, I will help you with your homework. No, I didn't watch the game. And about ice cream, maybe some other time."
	\end{quotation}
	Conversations of the uninvolved parenting style generated by GPT-4 generally had good quality with and without few-shot prompting. However, those generated with few-shot prompting demonstrated shorter lengths and followed the prompts even more closely.

\section{Discussion}
\label{sec:discussion}

Our study introduces a novel approach for conducting psychological and linguistic research through the use of simulated conversations. This methodology offers significant advantages, particularly for simulating human interactions involving potential unethical language or behaviors, which would not be permissible in traditional psychological experiments. For instance, researchers can now study scenarios such as harsh parenting conversations, which are ethically unsuitable for human participants.

The methods outlined in our research provide considerable flexibility in manipulating various aspects of the conversational agents. One could change the personality traits of the agents, the length and duration of the conversations, as well as the topics and language styles employed. Theoretically, infinite number of conversation exchanges could be produced. Such features are likely to be beneficial for future research focused on family dynamics and would like to implement nuanced controls, for example, inducing more agents or having controlled discussion topics. Moreover, the simulations can involve agents powered by different LLMs, facilitating a comparative analysis between them. For example, one could configure Agent A to operate based on a fine-tuned Mixtral model, while Agent B operates on a GPT-4 model. This capability opens up new avenues for future studies to create simulations with customized settings, enabling a more diversed exploration of linguistic and psychological phenomena.

The utilization of open-source LLMs holds substantial promise for conducting simulation research locally at minimal cost. Our findings indicate that the Mixtral-8x7b is conducive to such research, particularly when appropriate prompting techniques are employed. We anticipate that future iterations of local models will exhibit improved capabilities in managing simulations.

Our simulations, however, revealed issues such as repetitive conversations and inconsistencies in the personalities that did not align with the intended parenting styles. This suggests that while the approach is promising, refinement in the simulation design and agents' personalities are necessary. On the other hand, the quality of the conversations remains heavily influenced by the performance of the models themselves. For instance, conversations generated by the Mixtral-8x7b model frequently revolved topics about mathematics homework, a tendency we attribute to both the nature of the training data and the abstractness of the prompts used.

Therefore, we posit that models fine-tuned with specific example conversations could significantly enhance both the quality of simulations and the consistency of the data produced. We demonstrated the concept in Figure 3. If high-quality examples are used, the personalities of the agents could be rendered more distinctly, offering a clearer contrast to the relatively abstract psychological concepts currently employed in prompting. This approach would not only refine the simulation's fidelity but also its applicability in exploring complex psychological and linguistic phenomena.

\section{Conclusion}
The current study demonstrated a framework that utilizes LLMs for simulating family conversations involving four parenting styles. The simulations demonstrated great potential for conducting linguistic and psychological research that requires manipulating personalities, language styles, and mitigating potential ethical concerns with real human participants.

However, our simulations employed LLMs designed for general purposes, during which we observed instances of repetitive conversations and the emergence of non-conversational elements. These limitations highlight the challenges of using broadly trained models for specific simulation tasks. Nonetheless, these challenges can be mitigated through employing a few-shot prompting approach, fine-tuning models to cater to specific simulation requirements, or developing and applying more advanced models in the future.

{\it The work described in this paper was supported by the Start-up Fund for RAPs under the Strategic Hiring Scheme (1-BD5Y) provided by the Hong Kong Polytechnic University.}

\begin{figure}
	\centering
	\setlength{\fboxrule}{0pt}
	\setlength{\fboxsep}{0cm}
	\fbox{\includegraphics[width=\linewidth]{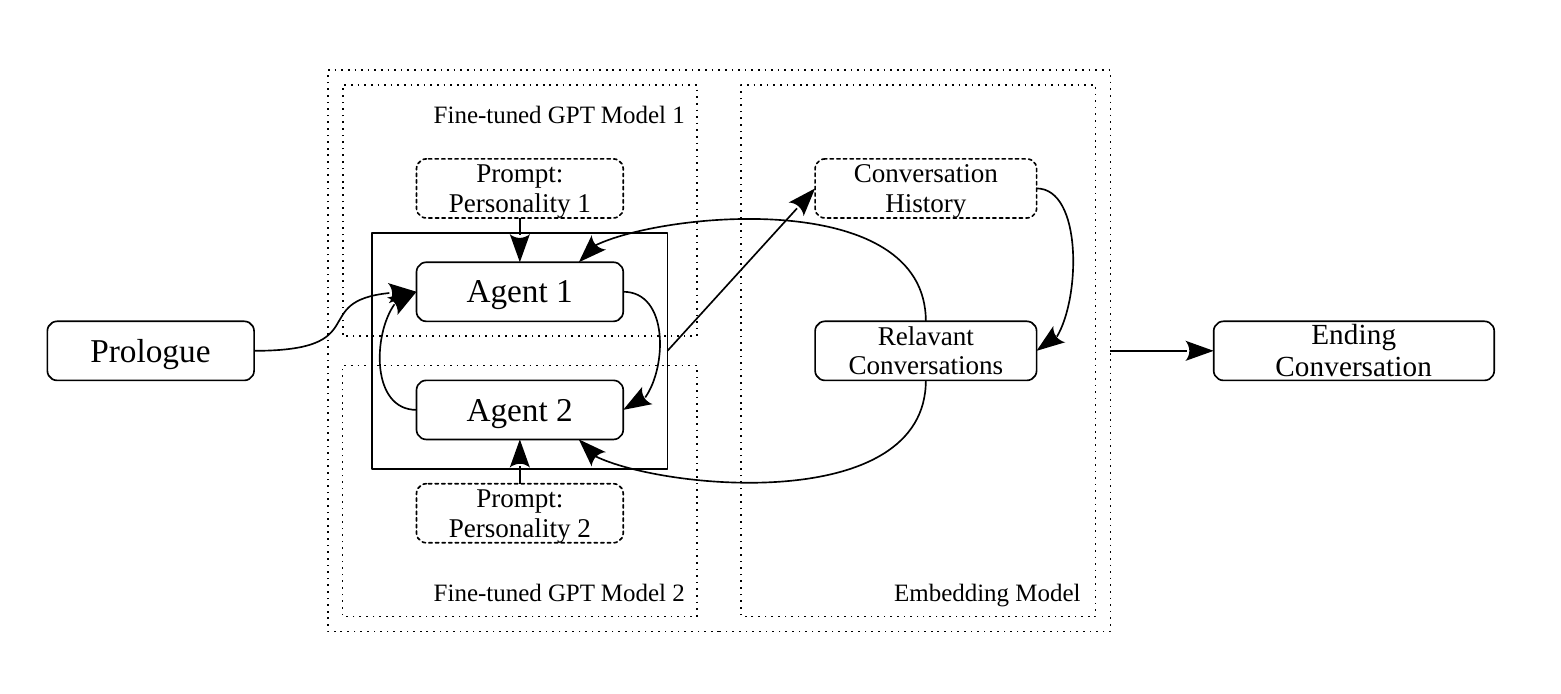}}
	\caption{Fine-tuned models may be used for better personality portray}
	\label{fig:fig3}
\end{figure}

\begin{figure}
	\centering
	\includegraphics[width=0.85\linewidth]{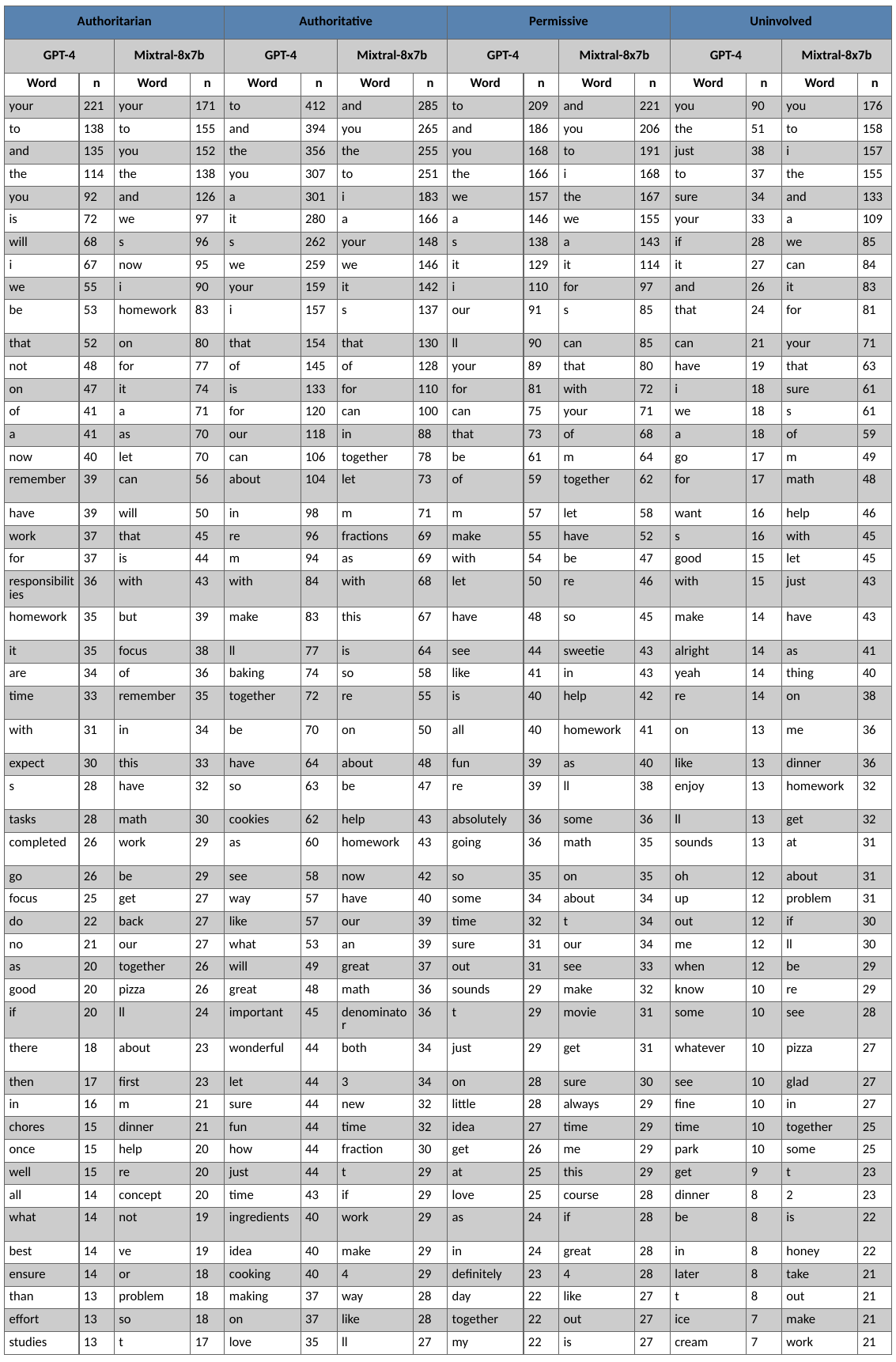}
	\caption{Word frequencies of generated conversations}
	\label{fig:tlb2}
\end{figure}

\bibliographystyle{unsrtnat}
\bibliography{references}

\end{document}